\documentclass[conference]{IEEEtran}
\IEEEoverridecommandlockouts
\usepackage{cite}
\usepackage{amsmath,amssymb,amsfonts}
\usepackage{graphicx}
\usepackage{graphics}
\usepackage{textcomp}
\usepackage{listings}
\usepackage[nohyperlinks, printonlyused, withpage, nolist]{acronym}
\usepackage[table,xcdraw]{xcolor}
\usepackage[noend]{algpseudocode}
\usepackage[T1]{fontenc}
\usepackage{lmodern}
\usepackage{pdfpages}
\usepackage{subcaption}
\usepackage{float}
\usepackage{hyperref}
\usepackage{color}
\usepackage{multirow}
\usepackage{hhline}
\usepackage{array}
\usepackage{enumitem}
\usepackage{tikz}
\usetikzlibrary{patterns}

\usepackage{pgfplots}
\pgfplotsset{compat=1.15}

\usepackage{tabularx}                       
\newcolumntype{L}[1]{>{\raggedright\arraybackslash}p{#1}}   
\newcolumntype{C}[1]{>{\centering\arraybackslash}p{#1}}     
\newcolumntype{R}[1]{>{\raggedleft\arraybackslash}p{#1}}    

\def\BibTeX{{\rm B\kern-.05em{\sc i\kern-.025em b}\kern-.08em
    T\kern-.1667em\lower.7ex\hbox{E}\kern-.125emX}}

\usepackage{setspace}

\newcommand{\CopyrightNotice}{\hbox%
    {\parbox{18cm}{\textsf\centering\scriptsize$ $\\[-11.5cm]
    \centering\setstretch{.3}%
    \textcopyright~2020 IEEE.
    Personal use of this material is permitted.
    Permission from IEEE must be obtained for all other uses, in any current or future media, including reprinting/republishing this material for advertising or promotional purposes,creating new collective works, for resale or redistribution to servers or lists, or reuse of any copyrighted component of this work in other works.\par}%
    \vspace{-\baselineskip}}%
}

\begin{document}

\begin{acronym}[WC]
\acro{wc}[WC]{WarpCore}
\acro{aos}[AOS]{{Array of Structs}}
\acro{soa}[SOA]{Struct of Arrays}
\acro{oa}[OA]{Open Addressing}
\acro{sc}[SC]{Separate Chaining}
\acro{lp}[LP]{Linear Probing}
\acro{qp}[QP]{Quadratic Probing}
\acro{dh}[DH]{Double Hashing}
\acro{cas}[CAS]{{\em compare-and-swap}}
\acro{cops}[COPS]{Cooperative Probing Scheme}
\acro{sm}[SM]{Streaming Multiprocessor}
\acro{simt}[SIMT]{Single Instruction Multiple Threads}
\acro{cg}[CG]{Cooperative Group}
\acro{tbb}[TBB]{Thread Building Blocks}
\acro{pram}[PRAM]{Parallel Random Access Machine}


\end{acronym}

\title{WarpCore: A Library for fast Hash Tables on GPUs
}

\author{
\IEEEauthorblockN{
    Daniel Jünger\IEEEauthorrefmark{1},
    Robin Kobus\IEEEauthorrefmark{1},
    Andr\'{e} Müller\IEEEauthorrefmark{1},
    Christian Hundt\IEEEauthorrefmark{2},
    Kai Xu\IEEEauthorrefmark{3},
    Weiguo Liu\IEEEauthorrefmark{3},
    Bertil Schmidt\IEEEauthorrefmark{1}
}
\IEEEauthorblockA{\IEEEauthorrefmark{1}
    Institute of Computer Science, {Johannes Gutenberg University}, Mainz, Germany \\
    Email: \{juenger, kobus, muellan, bertil.schmidt\}@uni-mainz.de
}
\IEEEauthorblockA{\IEEEauthorrefmark{2}
    NVIDIA AI Technology Center, Luxembourg, Luxembourg,
    Email: chundt@nvidia.com
}
\IEEEauthorblockA{\IEEEauthorrefmark{3}
    {School of Software}, {Shandong University}, Jinan, China,
    Email: \{xukai16@mail., weiguo.liu@\}sdu.edu.cn}
}

\maketitle
\CopyrightNotice

\begin{abstract}
Hash tables are ubiquitous. 
Properties such as an amortized constant time complexity for insertion and querying as well as a compact memory layout make them versatile associative data structures with manifold applications.
 The rapidly growing amount of data emerging in many fields motivated the need for accelerated hash tables designed for modern parallel architectures. 
In this work, we exploit the fast memory interface of modern GPUs together with a parallel hashing scheme tailored to improve global memory access patterns, to design WarpCore -- a versatile library of hash table data structures. 
Unique device-sided operations allow for building high performance data processing pipelines entirely on the GPU. 
Our implementation achieves up to 1.6 billion inserts and up to 4.3 billion retrievals per second on a single GV100 GPU thereby outperforming the state-of-the-art solutions cuDPP, SlabHash, and NVIDIA RAPIDS cuDF. 
This performance advantage becomes even more pronounced for high load factors of over $90\%$.
To overcome the memory limitation of a single GPU, we scale our approach over a dense NVLink topology which gives us close-to-optimal weak scaling on DGX servers.
We further show how WarpCore can be used for accelerating a real world bioinformatics application (metagenomic classification) with speedups of over two orders-of-magnitude against state-of-the-art CPU-based solutions. We plan to make our library publicly available upon acceptance of the paper.

\end{abstract}


\section{Introduction}
Hash tables are frequently used for storing key-value pairs in-memory because of their compact data layout and expected constant time complexity for insertion and retrieval. They are key data structures for 
bioinformatics \cite{pan2018optimizing}, computational geometry \cite{bisson2017high}, 
and deep learning \cite{chen2019slide}. This motivates the need for developing optimized implementations to support hash tables on modern architectures.

Common CPU-based hash table implementations such as \texttt{tbb::concurrent\_hash\_map} from the \ac{tbb} library or \texttt{std::unordered\_map} from the C++ standard library suffer from poor throughput induced by highly irregular memory access patterns during probing.
State-of-the-art accelerators may overcome this limitation by virtue of their fast high bandwidth memory (HBM2) and massive parallelism.

Consequently, a number of approaches have been designed for GPU-accelerated hashing using various probing schemes and memory access techniques. 
cuDPP~\cite{Alcantara2009, Alcantara2011}, Garcia et al. ~\cite{Garcia2011}, and StadiumHash \cite{Khorasani2015} were among the first to investigate hash map construction on GPUs proposing static implementations (i.e., no pairs can be added/deleted to an already constructed table) using one thread per key-value pair. 
More recent approaches including cuDF~\cite{cudf} (part of NVIDIA's RAPIDS framework), SlabHash~\cite{ashkiani2018dynamic}, and HashGraph \cite{Green2019HashGraph} are more flexible but are often limited in terms of performance or memory overhead.


We propose \ac{wc}, a highly efficient yet flexible library of hash data structures and algorithms that can achieve high performance for a variety of use cases.  
Our approach can achieve robust and often superior runtime performance even for very high load factors and storage densities.
The probing scheme is based on our previous WarpDrive method \cite{junger2018warpdrive} but eliminates its limitation to 32-bit single-value hash-tables. 
This is achieved by introducing a number of novel GPU-based data structures and associated algorithms within a versatile framework. Our detailed contributions are:
\begin{enumerate}
    \item The design of 32-bit and 64-bit massively-parallel single-value and multi-value hash table implementations with associated insertion/retrieval/deletion algorithms that allow for the flexible exchange of underlying data layouts.
    \item Host-sided and device-sided interfaces which enable high-throughput batch operations as well as concurrent processing of individual elements inside CUDA kernels. 
    \item We propose a novel memory-compact bucket list hash table with an associated dynamic growth scheme.
    \item We present techniques for concurrent execution of hash table operations and for the efficient usage of multiple GPUs.
    \item We show how \ac{wc} can be used for bioinformatics (metagenomic classification). 
\end{enumerate}

The rest of this paper is organized as follows. Section~\ref{sec:bg} provides some necessary background information. Related work is reviewed in Section \ref{sec:rw}. The design of \ac{wc} is presented in Section~\ref{sec:implementation}. Performance is evaluated in Section~\ref{sec:experiments}. Section~\ref{sec:conclusion} concludes the paper.

\section{Background}\label{sec:bg}

Hash maps are a class of data structures, that given a key $k$ from a sparse domain $K$, enable constant-time lookup of value $v \in V$ associated with that key thereby modelling functional dependencies $f:\, K \rightarrow V\,,\, k \mapsto f(k):=v$.
They avoid the memory overhead associated with dense look-up tables which hold memory for values associated with every possible key $k \in K$ by using a hash function $h: K \rightarrow I\,,\, k \mapsto h(k):=i$, mapping each key to a distinct memory index $i \in I$.

The complete set of keys $K$ is usually not known in advance which precludes the construction of a bijective mapping between $K$ and $I$, e.g., by using \emph{minimal perfect hash functions}.
For that reason and also due to performance considerations, a hash function $h$ is usually chosen to be non-injective thereby introducing potential index collisions $h(k)=h(k')$ for two distinct keys $k,k' \in K$. 
The most prevalent strategies for resolving such hash collisions are \ac{sc} and \ac{oa}.

\ac{sc} stores keys that map to the same hash $h(k)=i$ in a data structure tied to index $i$. This can either be a fixed array, a dynamic array, a linked list of contiguous chunks or a linked list of single elements. 
However, chaining shows several characteristics that are undesirable in the context of parallelization.
Linked lists usually involve cache-inefficient random access and require extra memory for pointers while using fixed-size arrays may lead to substantial memory over-subscription due to unused slots. Furthermore, lock-free insertion and deletion of nodes in linked lists can be error-prone due to pitfalls like the ABA problem and priority inversion.
 
With \ac{oa} colliding keys are stored in select locations taken from a sequence of candidate positions which are computed by a deterministic probing scheme. 
This approach is in general better suited for realizing efficient, lock-free updates and also for reasoning about their correctness. 
It is therefore often preferred for implementing concurrent hash tables. 
We also opted for \ac{oa} as hash conflict resolution technique for the same reasons.

A deterministic probing scheme generates a sequence $s(k,l)$ of candidate positions for storing a key $k$, where $l$ denotes the number of probing attempts.
Probing starts at the initial position $s(k, 0) = h(k)$ and continues as long as the candidate position is already occupied by another key or some abort criterion is met (e.g., all slots of the table have been visited).
The probing sequences of three prevalent schemes can be given as follows ($c = |I|$ denotes the capacity of the hash table):
\begin{itemize}
    \item \ac{lp}:
    $s(k, l) = \bigl(h(k)+l\bigr) \,\text{mod}\, c$ 
    \item \ac{qp}: $s(k, l) = \bigl(h(k)+l^2\bigr) \,\text{mod}\, c$ 
    \item \ac{dh}: $s(k, l) = \bigl(h(k)+l\cdot g(k)\bigr) \,\text{mod}\, c$ 
\end{itemize}
While \ac{lp} is cache-efficient, it tends to produce densely occupied regions that lead to a high variance in required probing length per key. 
This becomes especially pronounced when the number of inserted elements $n$ approaches the hash map capacity, i.e., its {\em load factor} $\alpha = \tfrac{n}{c}$ is high.
\ac{qp} and \ac{dh} avoid this so-called {\em primary clustering} using larger step sizes at the cost of more cache misses.
Extensions of these probing schemes have been proposed, among others Cuckoo Hashing \cite{Alcantara2009} and Robin Hood Hashing \cite{Celis1985}.

Fully featured CPU-based hash map implementations such as \texttt{std::unordered\_map} from the C++11 standard library support on-demand resizing in case the number of inserted elements $n$ exceeds the capacity $c$. A common strategy is to reinsert all data into a new hash map instance if the load factor reaches a critical threshold, e.g. $\alpha > 90\%$. These implementations typically also allow for the insertion of keys and values of arbitrary sizes. 
In case of concurrent table updates, modifications of each individual slot have to be serialized either by locking them with slow global mutexes or more efficient \ac{cas} operations.
We focus on the latter.
While this allows to issue concurrent inserts and queries without violating the integrity of the hash map, the outcome may depend on the actual execution order of the operations. 


While x86\_64 CPUs support \ac{cas} instructions for up to 128 consecutively stored bits, CUDA-enabled devices are limited to 64-bit words. Thus, packing key-value pairs $(k, v)$ into 64~bits enables the efficient use of \ac{cas} operations on an \ac{aos} memory layout. For larger keys and values, one can limit the \ac{cas} to the key slot of the struct or alternatively store keys $K$ and values $V$ separately as \ac{soa}. These variants use relaxed reads and writes to the value slots which might introduce priority inversion in case of simultaneously inserting distinct values for the same key. Whereas an \ac{aos} layout provides relatively high cache locality if both key and value of a slot are accessed, the effect reverses if we only access the key of each slot. 
In this case the values stored next to each key reduce the effective cache line size. This is especially critical if the value type is large compared to the key type. 

\section{Related Work}\label{sec:rw}



Several data-parallel GPU hash table implementations have been proposed which aim to leverage the fast memory bandwidth provided by modern GPUs.
Lessley et al.~\cite{lessley2019data} provide a comprehensive survey of these approaches and highlight the respective concepts and techniques used.

Alcantara et al.~\cite{Alcantara2009} were among the first GPU hash table implementations as part of the cuDPP library.
Their initial approach employs a two-stage table construction where keys are initially hashed into buckets of equal size residing in global memory.
Collisions are resolved with a third degree cuckoo hashing scheme.
Subsequently, the same authors proposed a single-pass variant~\cite{Alcantara2011} based on fourth degree cuckoo hashing which supports load factors of roughly $80\%$ achieving an insertion performance of up to 250 million inserts per seconds on a GTX 470.
cuDPP is limited to 32-bit wide types for both key and value. 
Also, tables are static, i.e., adding new key-value pairs to an already constructed table requires rebuilding the whole data structure.

CoherentHash~\cite{Garcia2011} introduces a data-parallel implementation of an \ac{oa} single-value hash table using Robin Hood hashing by augmenting each key with an additional 4-bit age indicator which trades the additional memory overhead with a lower on-average probing length.
It uses one thread for the lock-free insertion of a key-value pair using atomic \ac{cas} intrinsics and achieves comparable speed to cuDPP.

StadiumHash~\cite{Khorasani2015} employs an \ac{oa} strategy where the hash table itself may either reside in the GPU's global memory or inside host memory.
A so-called \textit{ticket board} residing in video memory is used to track slot occupation. It maintains a single bit indicating the slot's availability together with a small number of optional bits used as a signature of the key stored inside the slot.
If the full hash map can be kept in GPU global memory the performance of StadiumHash is between $1.04$x to $1.19$x faster than cuDPP on a GTX780 GPU at an average load factor of $80\%$.
In the case that the hash table is stored in host memory, i.e., out-of-core, the performance drops to around 100 million inserts per second restricted by the PCIe interconnect.
In order to better fit the GPU's SIMT execution model, StadiumHash employs a warp-cooperative work-sharing strategy, utilizing idling threads to cooperate in queued table operations.
Since only the ticket board has to be updated atomically, StadiumHash technically imposes no restrictions on the respective data types for keys and values.
However, the auxiliary ticket board implies additional memory overhead as well as additional random memory accesses per operation.
Phase-concurrent operations are guarded via exceptionally slow global device- or even system-wide barriers.

cuDF~\cite{cudf} is part of NVIDIA's RAPIDS framework~\cite{rapids} for manipulating columnar data frames on CUDA-enabled accelerators and also provides a data-parallel hash table implementations.
Similar to cuDPP, table construction is static and does not allow for subsequent or phase-concurrent insertions of new key-value pairs.
To the best of our knowledge, cuDF employs the only available multi-value GPU hash table.
However, their chosen linear probing scheme suffers from primary and secondary clustering effects for input distributions featuring many values per key, degrading performance significantly for these cases.

SlabHash~\cite{ashkiani2018dynamic} introduces a dynamic GPU hash table which follows the concept of \ac{sc} as its collision resolution strategy.
The table consists of an array of linked lists, each of which represents a chain of equally sized memory units, called \textit{slabs}, that store collided keys during insertion.
Each slab has a size roughly corresponding to that of a single cache line (128 bytes) and consists of multiple consecutive key-value slots and a single pointer to its successor slab.
SlabHash provides bulk operations which are executed using a warp-cooperative work-sharing strategy, where each CUDA thread inside a warp is assigned a distinct table operation such as insertion, retrieval, or deletion, which are then, one-by-one, executed cooperatively by all lanes inside the warp, ensuring memory coalescing.
Dynamic slab allocation during execution is realized through a specialized memory pool.
For bulk operations, they report $512$ ($937$) million operations per second for insertion (retrieval) on a Tesla K40c GPU, respectively.
Compared to cuDPP, SlabHash consistently achieves a lower throughput of insertions per second.
SlabHash achieves higher query throughput only when the average number of slabs per list is less than one.
Over all configurations, cuDPP attains the better query throughput.
However, on a newer Tesla V100 GPU, they consistently outperform cuDPP.
As stated in the corresponding manuscript, SlabHash supports single-value, as well as multi-value scenarios.
However, we found that the building blocks provided by the corresponding code repository~\cite{slabhashrepo}, were not sufficient to implement a multi-value retrieve operation.


HashGraph~\cite{Green2019HashGraph} handles hash-collisions with neither \ac{oa} nor \ac{sc}, but proposes a table construction method that is highly similar to a compressed sparse row matrix layout. HashGraph currently only supports static table builds, which again implies a lack of support for phase-concurrent workflows. Furthermore, the approach has high memory overhead since it requires $3n$ auxiliary memory during table construction with $n$ input key-value pairs.

Note that none of the above mentioned implementations feature out-of-the-box multi-GPU support.
With \ac{wc} we introduce a framework for constructing GPU hash tables that can overcome the aforementioned shortcomings of existing solutions while outperforming the state-of-the-art.

\section{Implementation}\label{sec:implementation}


 Our aim is the design of a versatile library for creating accelerated hash table data structures on CUDA-enabled GPUs. 
\ac{wc} provides optimized GPU implementations for the following data structures:
\begin{itemize}
    \item \texttt{HashSet}: stores set of keys; each key occurs only once 
    
    \item \textcolor{blue}{\texttt{SingleValueHashTable}}: stores key-value pairs; each key occurs only once
    
    \item \textcolor{blue}{\texttt{MultiValueHashTable}}: stores key-value pairs; same key may occur multiple times (with different values)
    
    \item \textcolor{blue}{\texttt{BucketListHashTable}}: stores all values associated with the same key in a linked list of buckets 
    
    \item \texttt{CountingHashTable}: counts distinct key occurrences
   
    \item \texttt{BloomFilter}: answers set membership queries
\end{itemize}

In this paper, we focus on the three highlighted types. The remaining types are built based on the same underlying principles. We now discuss some general library design features (\ref{sec:implementation:design}), single-value and multi-value hash table layout, parallel probing scheme, and associated operations (\ref{sec:implementation:oa}), our memory-compact bucket list (\ref{sec:implementation:bl}), concurrent execution (\ref{sec:implementation:concurrency}), and multi-GPU support (\ref{sec:implementation:multiGPU}),


\subsection{General Library Design Features} \label{sec:implementation:design}

\subsubsection{Modularity}

We provide building blocks that can be used to customize the basic data structures mentioned above and to create completely new one. Interchangeable parts include memory layout abstractions for switching between \ac{aos} and \ac{soa}, different hash functions and probing schemes. 

\subsubsection{Host-sided and Device-sided Interfaces}
The data structures in our library provide host-callable table operations which take input batches, enabling billions of independent table operations per second. 
We complement them with device-sided counterparts that work on single table elements.
This enables the building of pipelines where emitting key-value pairs, inserting them into a hash table and/or querying can be fused into monolithic CUDA kernels, avoiding costly global memory operations for intermediate results.
\subsection{Open Addressing Hash Table} \label{sec:implementation:oa}


\subsubsection{Memory Layout} \label{sec:implementation:oa:layout}
\begin{figure}[t]
    \centering
    \includegraphics[width=0.8\linewidth]{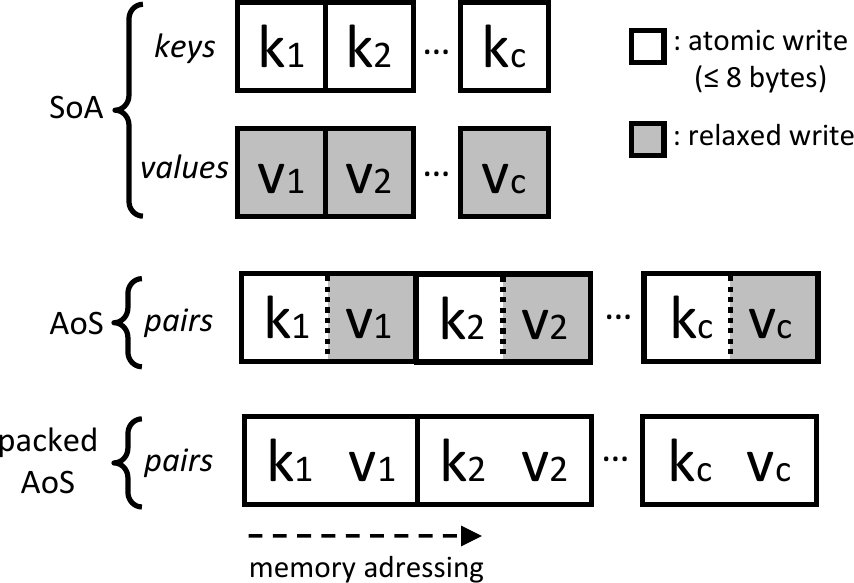}
    \caption[\ac{aos} versus \ac{soa} Data Layout]{\ac{soa}, \ac{aos}, and packed \ac{aos} memory layout of a key-value store with $c$ slots.}
	\label{fig:layout}
\end{figure}

Our basic \ac{oa} hash table consists of contiguous arrays residing in global GPU memory in either \ac{aos} or \ac{soa} layout, i.e., one array for holding aggregates composed of a key and a value member, or alternatively two arrays of the same length where the first holds the keys and the second holds the corresponding values. 
The array size determines the maximum number of key-value pairs (capacity) the hash table can hold.
We initialize each key slot with an empty-indicator $k_e$ in order to distinguish empty slots from occupied ones during probing.
For the case that both key and value data types do not exceed a width of 32~bits, we provide a packed \ac{aos} layout, where a key-value pair is bit-packed into a single 64-bit unsigned integer. 
This allows for storing both key and value by using a single atomic \ac{cas} operation instead of an initial atomic swap of the key into its target slot followed by a relaxed store of the value. Figure \ref{fig:layout} illustrates the three supported memory layouts.

\subsubsection{Parallel Probing Scheme} \label{sec:implementation:oa:cops}

A na\"ive approach assigns each key to its own CUDA thread. However, since each key typically follows a different probing sequence this would lead to  non-coalesced global memory accesses.
An alternative approach could use an entire warp of 32 threads per input key $k$, such that each thread with lane ID $t$ probes a different hash table position $h(k,t)\bmod c$.
If any thread finds a matching slot it can signal the other threads in the warp to terminate probing via fast register vote intrinsics.
However, this approach is only beneficial if the hash table positions $\{h(k,0)\bmod c, \ldots, h(k,31)\bmod c\}$ fall within the same memory region, enabling threads in a warp to share the same cache line. The only known probing scheme that meets this constraint is \ac{lp} which suffers from 
primary clustering. 

Thus, we rely on a hybrid approach, called \ac{cops} consisting of an inner intra-warp probing scheme combined with outer
probing based on \ac{dh}. The inner scheme is based on \ac{lp} and ensures data locality between threads inside a warp. The outer probing scheme determines the starting index offset for the inner scheme. The resulting local probing sequence of a warp for the $i$th probing attempt is $\{\left( h(k,\lfloor\frac{i}{32}\rfloor)+0\right)\bmod c, \ldots, \left( h(k,\lfloor\frac{i}{32}\rfloor)+31\right) \bmod c\}$. 
\ac{dh} features both low probing lengths and fairly uniform slot occupation patterns within the table. Additionally, we maintain \ac{dh}'s property of cycle-freeness by selecting $c = p\cdot32$, where $p$ is prime whilst ensuring that the second hash function 
generates step sizes as multiples of $32$. 

Assuming a uniform distribution of populated slots and a load factor of $90\%$, every tenth slot would be empty on average. Thus, it would be likely that an empty slot can be found by a group of less than $32$ threads at the first probing attempt. This motivates the usage of sub-warp tiling based on CUDA's \ac{cg} feature which enables us to use thread groups of sizes 1, 2, 4, 8, 16, or 32. 

\begin{figure}[t]
    \centering
    \includegraphics[width=\linewidth]{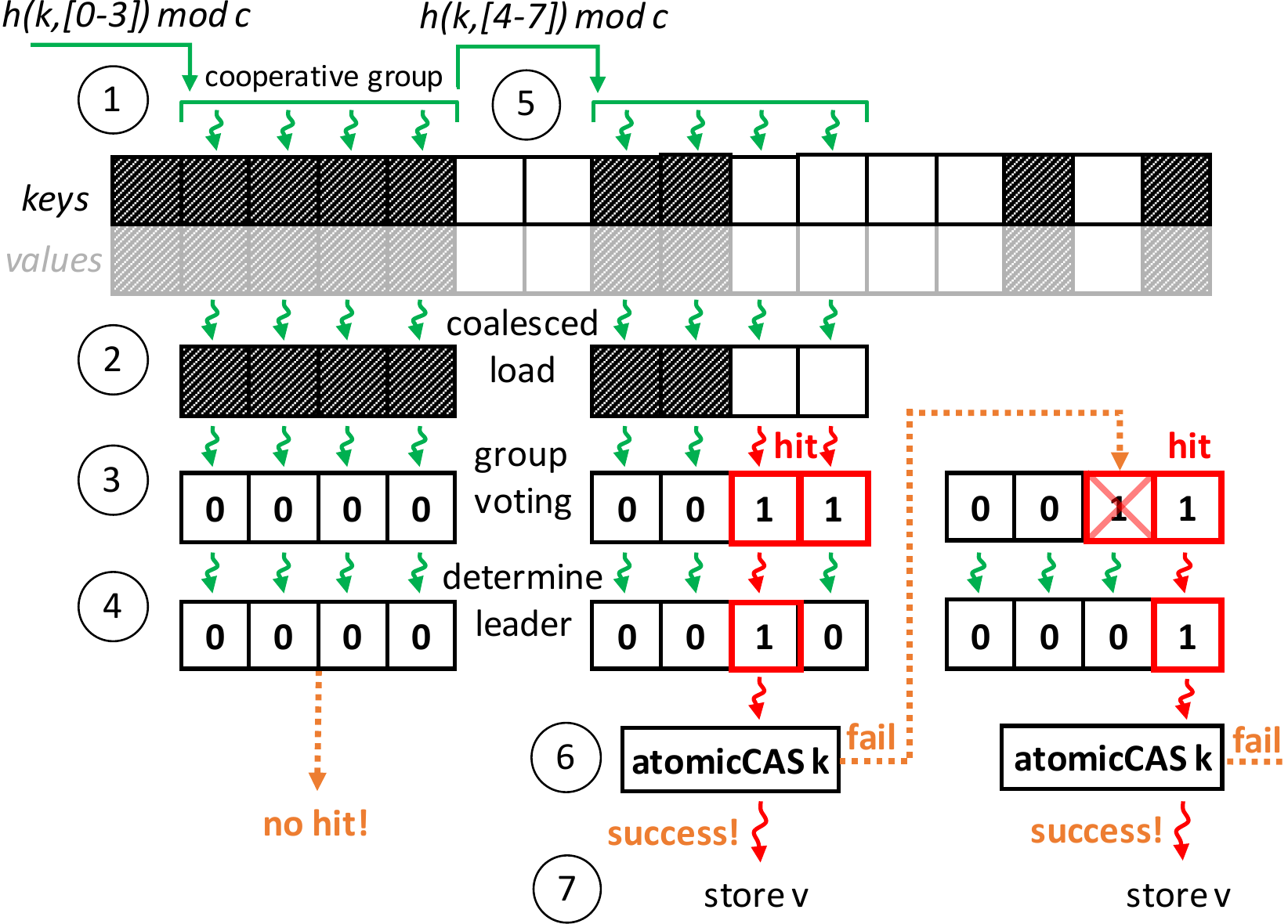}
    \caption[Cooperative Probing Scheme]{Insertion of a key-value pair $(k,v)$ into a hash table with capacity $c$ using \ac{cops} with an outer probing scheme $h$, an inner probing window size of 4 and a \ac{cg} size of 4.}
	\label{fig:insert}
\end{figure}

Figure~\ref{fig:insert} depicts the insertion of a key-value pair $(k,v)$ into a hash table
with 
a \ac{cg} size of 4 
based on seven steps:
\begin{enumerate}[label=(\arabic*)]
    \item The outer probing scheme is used to determine the initial probing index.
    
    \item Each thread in a \ac{cg} loads one key slot from global memory in coalesced fashion.
    
    \item Each thread checks whether its assigned slot is a potential candidate for inserting $(k,v)$ and communicates its result via a fast in-register group vote.
    
    \item  The thread associated with the lowest candidate slot index is selected.
    
    \item  If there are no candidate slots (left column in Figure~\ref{fig:insert})
    steps 1 to 4 are repeated until a suitable one is found.
    
    \item We try an atomic \ac{cas} of the key into the selected candidate slot. If it fails (due to a collision with a successful insertion by another thread), we repeat steps 4 and 6 until the \ac{cg} has no candidate slots left. In this case we start from Step 5. 
    
    \item If the key was inserted successfully in Step 6, a relaxed store operation is issued to write the value.
    
\end{enumerate}

Probing with variable CG sizes inside the same table can be accomplished by setting the inner probing window to a fixed size, e.g. 32. Smaller groups iterate over said window in a linear fashion before continuing with the outer probing scheme, which ensures probing consistency for all group sizes. 

\subsubsection{Insertion} \label{sec:implementation:oa:insertion}

Inserting a new key-value pair is accomplished by using \ac{cops} for both single-value and multi-value context. 
For a single-value hash table, an additional warning is issued if the key currently being probed for is already present.
While the device-sided function inserts a single key-value pair into the table and is executed inside a \ac{cg}, the host-callable batch operation consists of a data-parallel CUDA kernel. 
For a batch insertion of $n$ key-value pairs we start a kernel consisting of $n\cdot g$ many threads, where $g$ denotes the \ac{cg} size each query is executed in. 
Each \ac{cg} in the grid is assigned to a single pair and calls the corresponding device-sided insert function, implementing a data-parallel scheme over the input batch.

\subsubsection{Retrieval} \label{sec:implementation:oa:retrieval}

Retrieval relies on the same scheme but instead of probing for an empty slot, we look for slots that hold the queried key. Search can terminate when we encounter an empty slot before finding the queried key,
since \ac{cops} guarantees that any key is inserted at its lowest possible index in the probing sequence.

The device-sided retrieval function takes a query along with a \ac{cg} in which the operation should be executed. 
The host-sided function reads a batch of keys from global memory and writes the retrieved values to a user-allocated output array on the device.
In the single-value case, the number of returned values cannot exceed the number of queried keys and probing can stop after the first value of a query is found. 
In the multi-value case the number of values associated with a query is not known in advance.
Thus, if all retrieved values should be written to memory, the size of the output array has to be determined in a separate counting pass beforehand.
The retrieval of all values associated with a batch of $n$ queries requires an additional temporary array of size $n+1$ storing the offsets of the per-key value segments in the output array.
The offset array is computed using a prefix sum over the number of values per query. 

In some cases it is sufficient to process the values associated with each key one-at-a-time on the device instead of copying the retrieved values of a host-sided batch query into a distinct new location in global device memory.  
\ac{wc}'s data structures provide higher-order member functions \texttt{for\_each(keys,callback)} and \texttt{for\_all(callback)} which take a device-sided callback function (object), e.g., a device-sided lambda function. 
The callback is invoked in parallel for each query found in the table during probing and receives the corresponding key-value pair and the key's index.

\subsubsection{Deletion} \label{sec:implementation:oa:deletion}

Deleting keys is accomplished by overwriting the table slot with a tombstone $k_t\neq k_e$.
During insertion, a slot with a tombstone is treated as a regular empty slot which can be re-populated with a new key-value pair. During retrieval, it is interpreted as a populated slot. 

\subsection{Bucket List Hash Table} \label{sec:implementation:bl}

Due to the relatively small amount of available video memory, effective memory utilization often plays a crucial role. We define the \emph{storage density} $\rho$ of a container data structure holding data elements, as the amount of stored information bits over the total amount of memory allocated by the container. For single-value \ac{oa} hash table designs $\rho$ is equivalent to the table's load factor $\alpha$, i.e. the number of occupied key-value slots over the number of allocated slots. However, if the same key occurs more than once it is stored multiple times, thereby degrading storage density.
For use cases where this is undesirable, like in Section \ref{sec:experiments:usecases:metacache}, we provide an alternative multi-value hash table which links all values to a single instance of the corresponding key. 


Our bucket list hash table uses a single-value hash table with a primary list handle as value. Actual values are stored in linked lists of contiguous memory chunks called buckets. 
The list handle is a 64-bit packed, atomically updatable data structure consisting of three fields: a pointer to the last list node associated with a key, the total number of values per key and 2~bits indicating one of four possible states (uninitialized, blocked, ready, full).
Transitions between states are guarded by atomic \ac{cas} operations on the primary list handle ensuring that list operations are linearizable. 

\begin{figure}[t]
    \centering
    \includegraphics[width=0.7\linewidth]{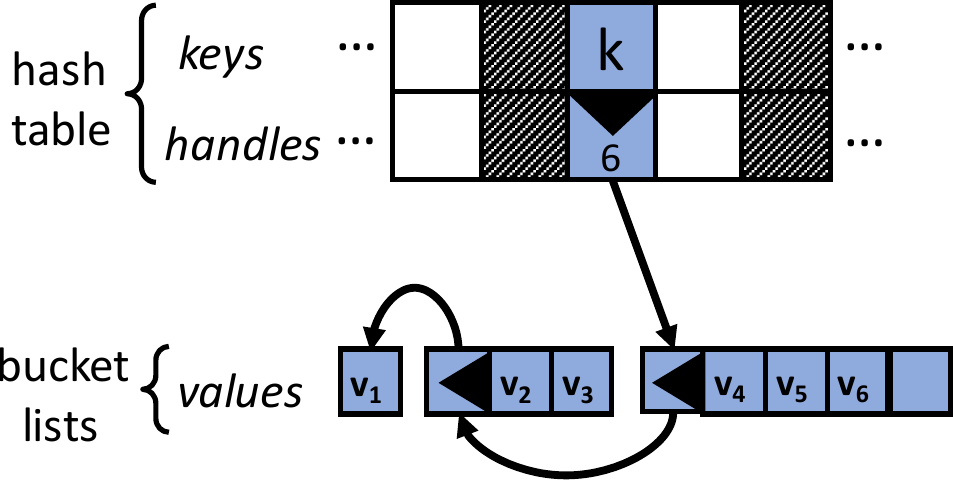}
    \caption[]{
    Example of a bucket list storing six values associated with key $k$. Keys are stored in a single-value hash table together with a handle to the associated bucket list that holds a reference to the last bucket and the total number of values inside the list.  
    }
	\label{fig:bucket_list}
\end{figure}

The leading slot of each but the first bucket consists of a reference that points to the previous bucket in the list, followed by the actual value slots. 
Since the exact distribution of values per key is usually unknown in advance, we propose the following growth strategy. When a key is inserted into the hash table for the first time, its initial value bucket of size $s_0$ is allocated.
Subsequent bucket sizes are set to $s_i = \lceil \lambda \cdot s_{i-1} \rceil$ where $\lambda \geq 1$ denotes the bucket growth factor.
Figure~\ref{fig:bucket_list} shows an example using growth parameters $s_0 = 1$ and $\lambda = 2$.
If the input data distribution is known both parameters can be used to adapt the growth strategy for improved memory efficiency.
Since global memory allocations would act as a device-wide barrier, memory for buckets is pre-allocated as a single contiguous array which serves as a memory pool.

Inserting a key-value pair starts with probing for the key in the single-value hash table. 
If the key is not already present, we insert it into the hash table and allocate its first value bucket.
One thread atomically marks the associated list handle as blocked, thereby ensuring that is has exclusive access to the value list, and subsequently requests a new bucket from the memory pool. 
If successful, the value is appended to the list and the list's handle is unblocked.
Other threads that encounter a blocked handle implement a busy waiting strategy with exponential back-off until the handle has been unblocked.
In case the key is already present in the table, we check if the current tail bucket has any empty value slots available.
If true, we try to reserve a slot inside the bucket by an atomic increment of the list handle's value counter.
If the atomic \ac{cas} operation is successful, we write the value to the reserved slot and terminate.
However, if this operation fails due to another thread successfully inserting a new value concurrently, we reload the altered handle from memory and repeat until insertion has succeeded. 
If the current tail bucket is full, we attempt to block the handle and request a new bucket from the memory pool.

\begin{figure}[t]
    \centering
    \includegraphics[width=\linewidth]{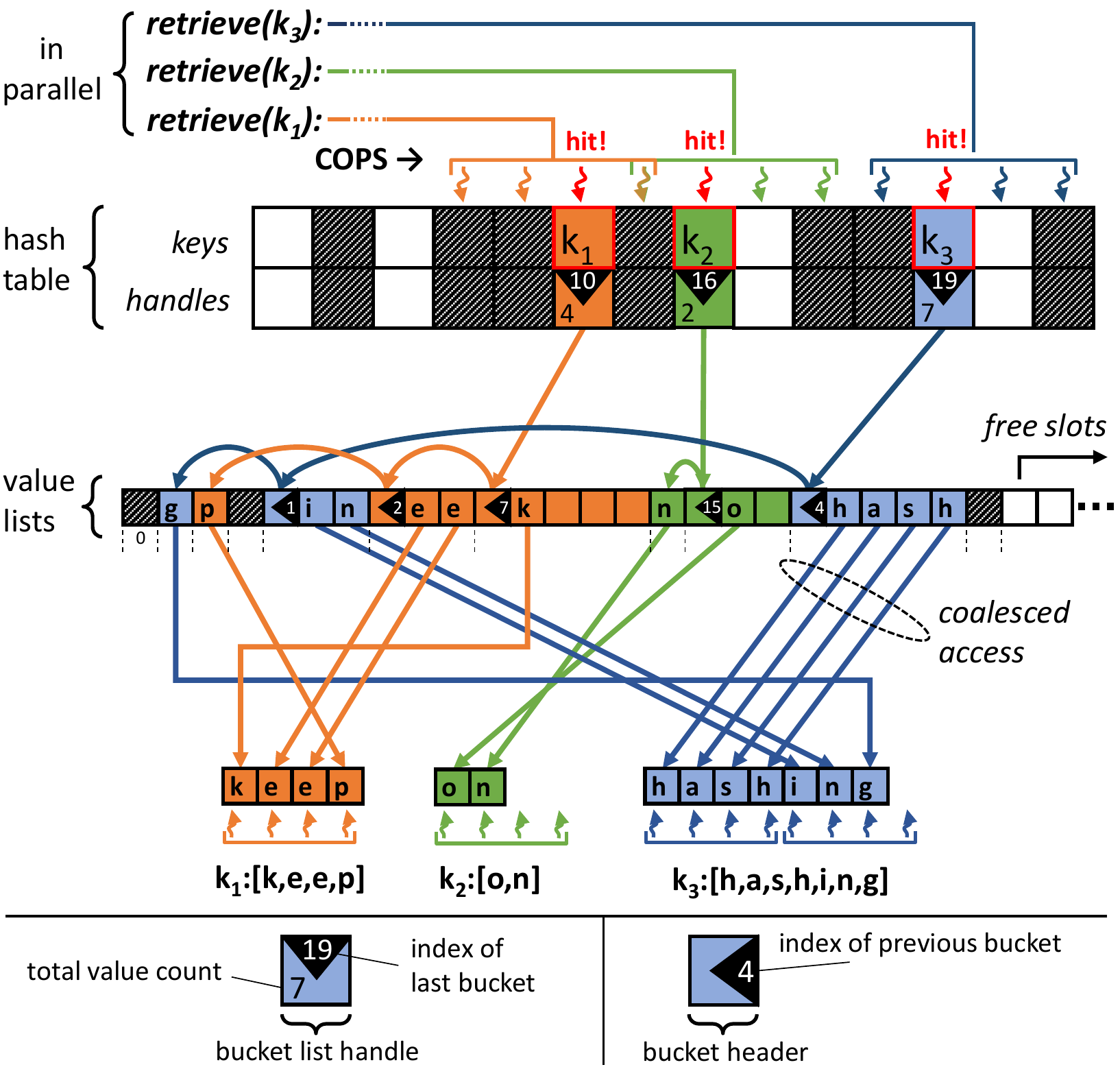}
    \caption[]{Example of retrieving all associated values for a batch of keys $\{k_1, k_2, k_3\}$ from the bucket list hash table.}
	\label{fig:bucket_list_retrieve}
\end{figure}

As shown in Figure~\ref{fig:bucket_list_retrieve} retrieval of a batch of keys is done by probing for each key in parallel.
If a key is found, threads in the current \ac{cg} begin following the bucket list references until reaching the initial bucket.
Values are read from the buckets using as many threads from the current \ac{cg} as possible, thereby enabling coalesced access for sufficiently large buckets. In case the \ac{cg} size exceeds the current bucket size, remaining threads diverge and proceed to the next bucket.

\subsection{Concurrency} \label{sec:implementation:concurrency}
All device-sided table operations can be overlapped and only synchronize with the \ac{cg} they are executed in. 
The majority of hash table operations in \ac{wc} can be executed asynchronously (with respect to both host CPU and GPU). 
Only operations that use function return values to return data back to the host block until the return parameter is available. 
By default, all operations are executed in the default stream of the GPU 
making them blocking function calls. 
However, when called with a non-default stream, operations are issued asynchronously with respect to other streams, CUDA-devices, as well as the host system. This is a crucial feature for employing multiple hash tables residing on different devices in a multi-GPU environment.

Note that not all operations on the same hash table can be safely executed concurrently.
We distinguish two categories of operations based on whether they modify the internal state of a table. 
Overlapping the same operation and all read operations is always valid. 
However, overlapping write operations with another operation of a different kind is not supported due to two concerns:

\begin{figure*}[!t]
    \centering
    \begin{subfigure}[b]{.47\linewidth}
        \includegraphics[width=\linewidth]{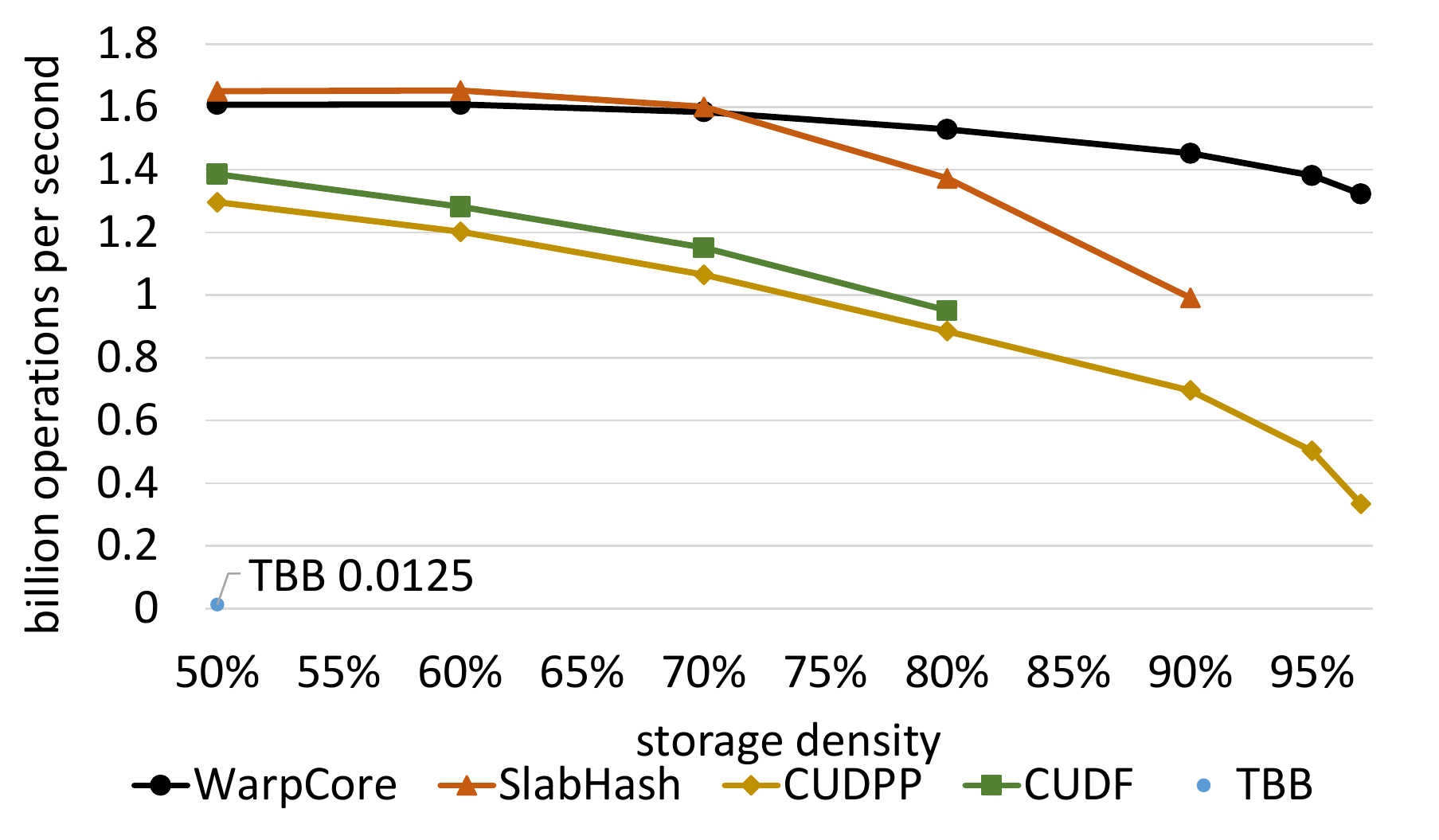}
        \caption[Single-value Hash Table Insertion Performance]{Bulk insertion performance.}
    	\label{fig:eval_single_value_insert}
    \end{subfigure}
    \hfill
    \begin{subfigure}[b]{.47\linewidth}
        \includegraphics[width=\linewidth]{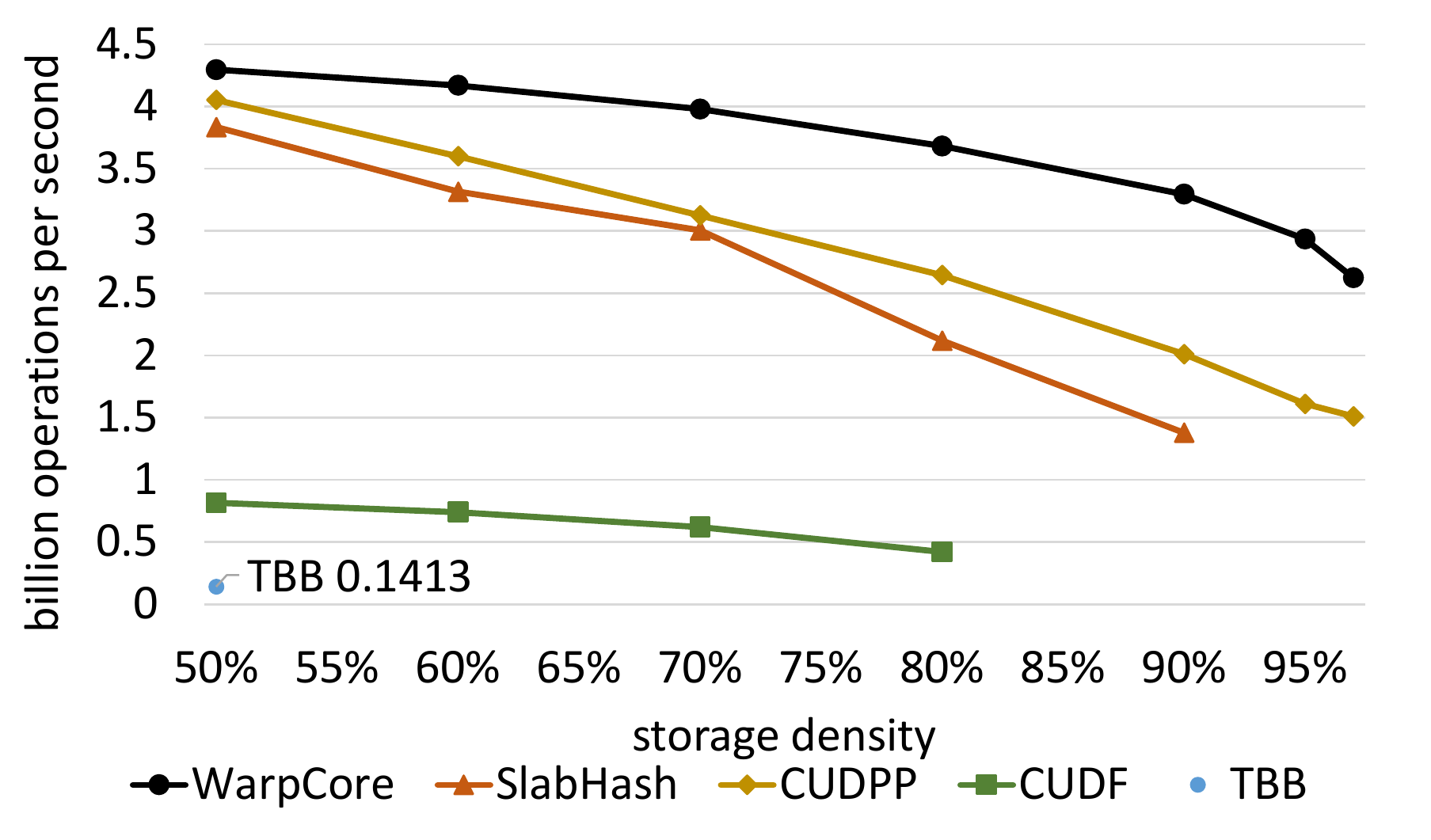}
        \caption[Single-value Hash Table Retrieval Performance]{Bulk retrieval performance.}
    	\label{fig:eval_single_value_retrieve}
	\end{subfigure}
	\caption[Single-value Hash Table Performance]{Performance comparison of different single-value hash table implementations during bulk operations for $2^{28}$ (2~GB) unique key-value pairs.}
	\label{fig:eval_single_value}
\end{figure*}

\begin{enumerate}[label={\color[HTML]{ab0808} (\arabic*)}]
    \item If the combined key-value pair type exceeds the size constraint of 64~bits for CUDA atomic operations, insertion of a new key-value pair consists of two memory operations, an atomic swap of the key followed by a relaxed write of the value. If insertion and retrieval of the same key were to occur simultaneously, reading the value might yield incorrect results.
    
    \item Overlapping insertions and deletions is prone to a variant of the ABA problem.
    If we simultaneously issue a deletion combined with two insertion operations, all working on the same key $k$ which is not yet present in the table, there exists a possible execution order in which a race condition may occur, leaving the table in an invalid state.
\end{enumerate}



If the hash table is configured to use 64-bit packed key-value pairs that can be stored using a single atomic operation, all possible combinations of operations leave the table in a valid state and return valid results.
Nevertheless, combining insertion with deletion or any write operation with any other read operation might still be undesirable if these operations work with the same keys concurrently. The final result may thus depend on their execution order and leave the table in an unpredictable, albeit valid, state.

\subsection{Multi-GPU Support} \label{sec:implementation:multiGPU}
The limited amount of main memory available on a single GPU can be insufficient for many data-intensive applications. Thus, \ac{wc} allows for building and querying data structures on multiple GPUs. There are two modes of operation: {\em distributed} and {\em independent}. 

The distributed mode assigns each key (and its associated values) to exactly one distinct GPU. This is done by first partitioning keys of an input batch according to their corresponding GPU ID by means of a device-sided multi-split \cite{Ashkiani2016} followed by scattering these segments to the GPUs where they belong.
In case each participating GPU holds a separate input batch, we use an all-to-all communication primitive on NVLink connected systems \cite{kobus2019gossip} to simultaneously exchange segments between all GPUs.
This approach has the advantage that (multi-value) retrieval does not require merging the results of individual GPUs, since each key may only reside on one GPU.

The second mode simply constructs and stores one independent hash table per GPU. This can be desirable in cases where result merging is acceptable or can be done without communicating all values. Data to be inserted is simply scattered and queries are broadcast to all GPUs.

\section{Experimental Evaluation} \label{sec:experiments}

Experiments were conducted on the following systems:
\begin{description}
\item[System~1:] Dual-socket Intel Xeon GOLD 6238 CPU (2x22 cores at 2.10 GHz) with 192 GB DDR4 RAM and 2 Quadro GV100 GPUs connected via NVLink each with 32 GB HBM2 memory running Ubuntu 18.04 LTS, CUDA 10.2, GCC 8.3.0. 
\item[System~2 (DGX-1):] Dual-socket Xeon E5-2698 v4 CPU (2x20 cores at 2.20 GHz) with 512 GB DDR4 RAM and 8 Tesla V100 GPUs connected by NVLink each with 32 GB HBM2 memory running Ubuntu 18.04, CUDA 10.1, GCC 8.3.0. 
\end{description} 

Time measurements are accomplished with CUDA event system timers. In all experiments, we assume that the data to be inserted or retrieved resides either in host RAM for operations executed on the CPU or in video RAM for device-sided benchmarks. 

\subsection{Single-Value Performance}\label{sec:experiments:single_value}

We evaluated our single-value hash table against the publicly available state-of-the-art GPU implementations cuDPP~\cite{Alcantara2011}, SlabHash \cite{ashkiani2018dynamic}, and cuDF \cite{cudf}. 
Additionally, we included a widely-used multi-threaded CPU implementation, namely \texttt{tbb::unordered\_hash\_map} from \ac{tbb}. 
Our benchmark scenario consists of an initial bulk build operation which inserts a set of $2^{28}$ unique 4-byte keys along with 4-byte arbitrary values into each hash table.
Subsequently, we query the same set of keys against the table and retrieve their corresponding values.
For both phases, we measure the number of executed operations per second averaged over ten consecutive runs in reference to the target density of the data structure, i.e., after all pairs have been inserted (see Figure~\ref{fig:eval_single_value}). 
Benchmarks were conducted on a single GV100 GPU on System~1, while \ac{tbb} utilizes all 44 CPU cores.
Note that the target storage density of \ac{tbb}'s hash table cannot be set by the user.
Regarding insertion performance, \ac{wc} outperforms cuDPP and cuDF by a factor of $3.95$ and $1.6$ for $\rho = 0.97$ and $\rho = 0.8$, respectively. 
For relatively low densities, SlabHash's performance is on par with \ac{wc}. 
However, if $\rho$ increases ($>70\%$), \ac{wc}'s cooperative probing scheme is superior.
Note that some implementations did not finish their execution for densities above a certain threshold.
For retrieval, \ac{wc} is faster than all competitors by a factor of $8.76$ ($\rho = 0.8$), $1.64$ ($\rho = 0.9$), $2.39$ ($\rho = 0.9$) compared to cuDF, cuDPP, and SlabHash, respectively. 
Our results show, that \ac{wc} outperforms all other tested GPU implementations at high storage densities and is also over two orders-of-magnitude faster than \ac{tbb}.

\begin{figure}[t]
    \centering
    \includegraphics[width=0.9\linewidth]{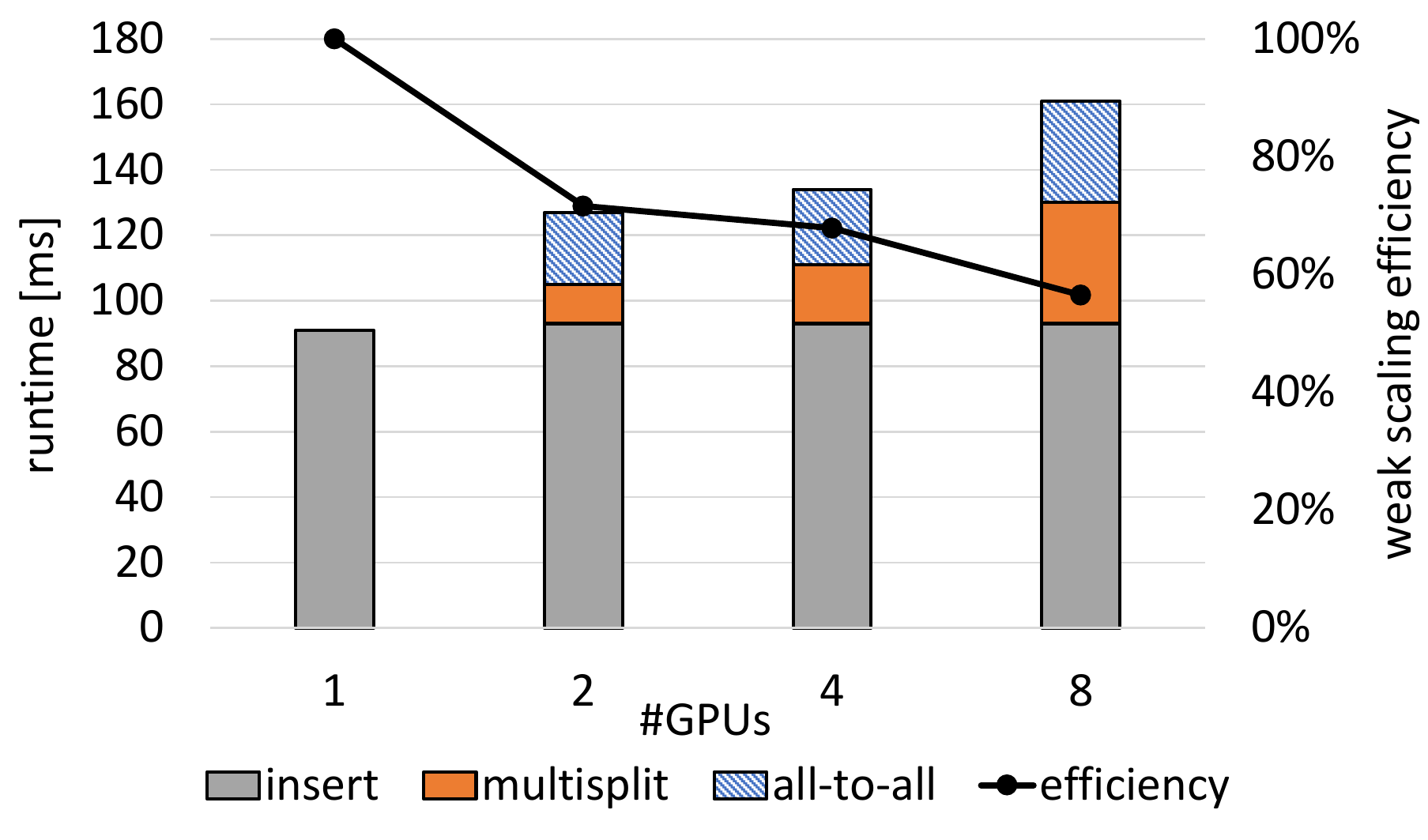}
    \caption[]{Weak scalability analysis of hash table insertion on System~2 (DGX-1) with 2~GB of input data per GPU.}
	\label{fig:scalability}
\end{figure}

\begin{figure*}[t]
    \centering
    \begin{subfigure}[b]{.47\linewidth}
        \includegraphics[width=\linewidth]{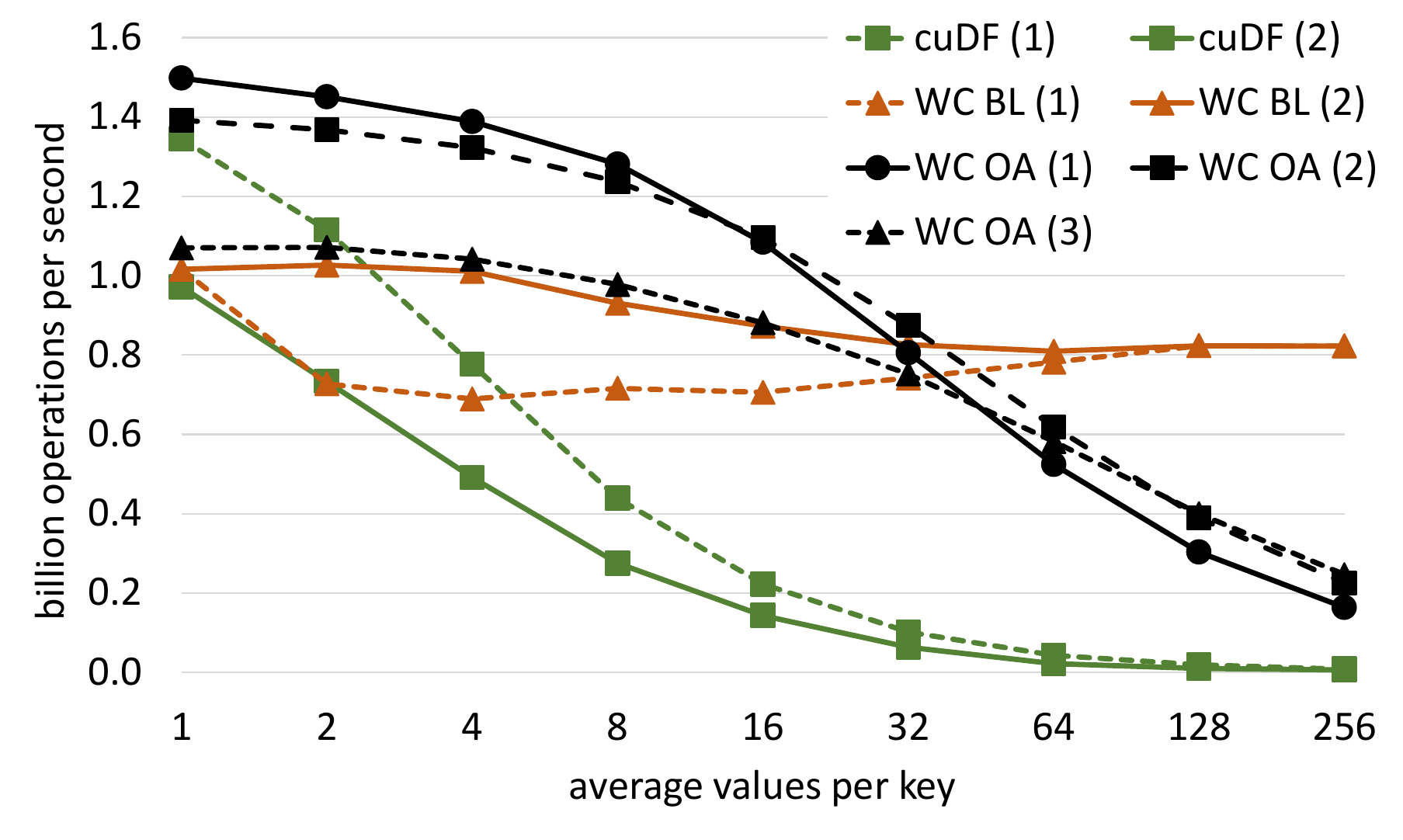}
        \caption[Multi-value Hash Table Insertion Performance]{Bulk insertion performance.}
    	\label{fig:eval_multi_value_insert}
    \end{subfigure}
    \hfill
    \begin{subfigure}[b]{.47\linewidth}
        \includegraphics[width=\linewidth]{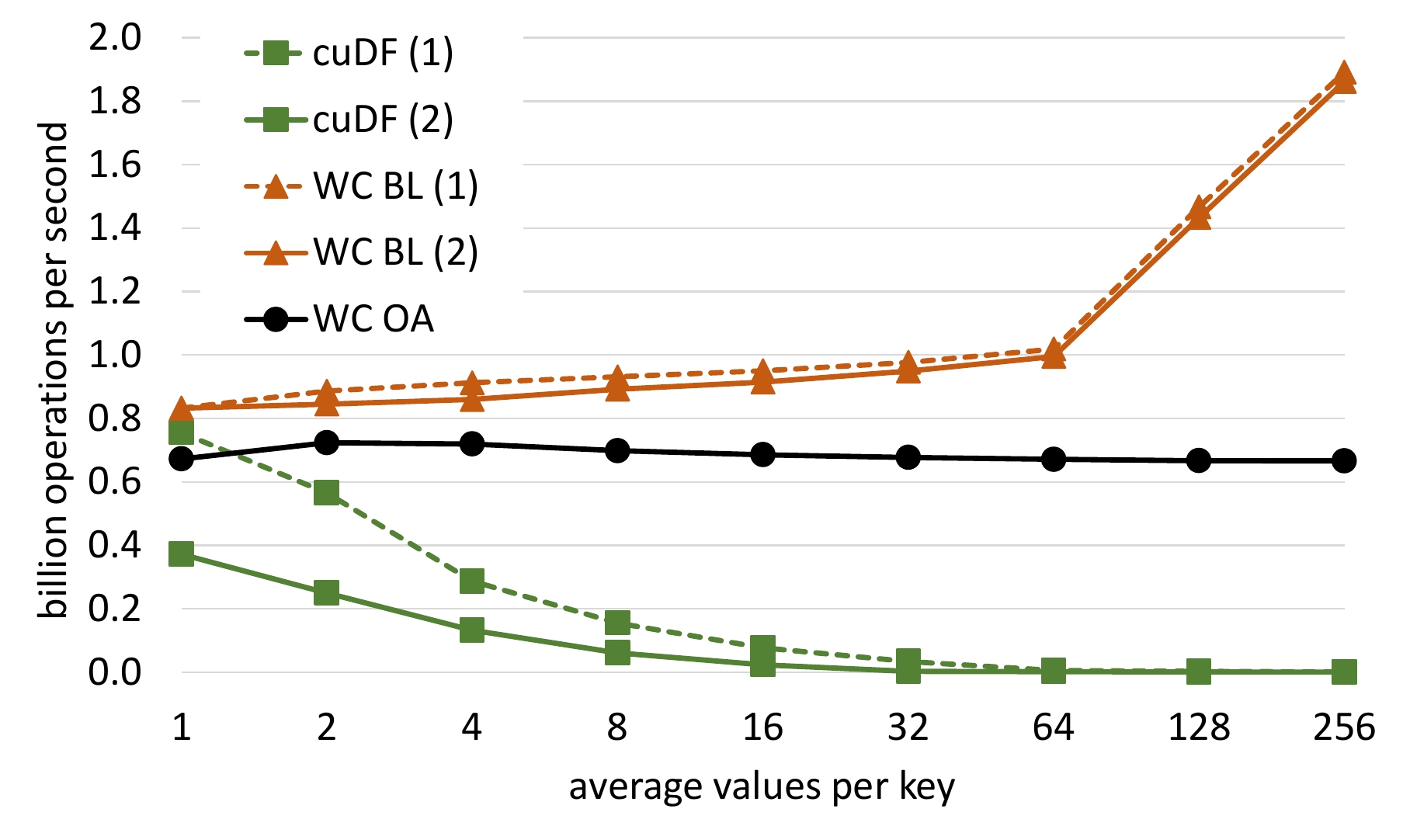}
        \caption[Multi-value Hash Table Retrieval Performance]{Bulk retrieval performance.}
    	\label{fig:eval_multi_value_retrieve}
	\end{subfigure}
	\caption[Multi-value Hash Table Performance]{
	Performance of multi-value hash table implementations in billion operations per second for bulk operations on $2^{28}$ (2~GB) key-value pairs with a varying number of values per key. We compare our \ac{wc} \ac{oa} variant with a target load factor of 0.8 for relevant CG sizes of 8 (1), 16 (2), and 32 (3), cuDF with two target load factors 0.5 (1) and 0.8 (2), as well as our bucket list variant (BL) with default ($\lambda=1.1$, $s_0=1$) (1) and optimal growth strategy ($\lambda=1.0$, $s_0 =$ average number of values per key) (2).
	}
	\label{fig:eval_multi_value}
\end{figure*}

To evaluate multi-GPU scalability, we tested the distributed mode discussed in Section~\ref{sec:implementation:multiGPU} on System~2 with $2^{27}$ unique 8-byte keys along with 8-byte values as input. 
Figure~\ref{fig:scalability} shows a weak scalability analysis with runtime breakdowns for data partitioning, communication, and insertion together with the achieved efficiency.
Note, that it would be possible to further apply common optimization strategies like batching and overlapping CUDA streams to hide the runtime of data transfers behind the kernel execution for multi-split and insertion, but we decided to report the full runtime of each primitive to show the relative cost.

\subsection{Multi-Value Performance}\label{sec:experiments:multi_value}

To conduct the multi-value benchmark, we control the average number of identical keys $r$ in the input batch by drawing $n$ elements uniformly random from the range $(1,\ldots,\frac{n}{r})$.
Figure~\ref{fig:eval_multi_value_insert} shows the results for inserting such a distribution of keys into different hash tables for varying values of $r$ and a fixed target load factor for \ac{wc} and cuDF (cuDPP and SlabHash are only designed for single values).
In case of our bucket list hash table, this load factor is solely enforced on the key store, i.e., the \ac{oa} hash table holding the keys along with bucket list handles.
During retrieval we probe for the complete range of $n$ unique keys $(1,\ldots,n)$. 
With $r$ increasing, this results in some of the queried keys to not being present in the table, whilst other keys are associated to multiple values.
Using this setup, the total number of retrieved values is always equal to the number of input queries, i.e. $n$, which eliminates any effects of I/O skew from our measurements.

For insertion, our \texttt{MultiValueHashTable} (\ac{wc} \ac{oa}) shows comparable performance to its single-value counterpart in the case that input keys are close-to unique.
When the value multiplicity increases, throughput degrades due to longer probing sequences.
cuDF shows the same behaviour but handles high key multiplicities worse due to its \ac{lp} scheme, which is prone to primary clustering.
This effect is amplified in a multi-value setup, where multiple identical keys collide in the same cluster of initial probing position.
\ac{wc} \ac{oa} consistently outperforms cuDF during insertion.
For value multiplicities $\leq 16$ a CG size of 8 shows optimal performance. However, for higher multiplicities larger CG sizes are more beneficial.
Both tested variants of \texttt{warpcore::BucketListHashTable} (\ac{wc} BL) are slower than \ac{wc} \ac{oa} if the average number of values per key is less than 32 but show nearly constant performance for higher multiplicity, while \ac{wc} \ac{oa} gradually degrades.
We suspect that this is a trade-off between \ac{wc} BL's additional steps required after probing, i.e., appending the value to the key's bucket list and \ac{wc} \ac{oa}'s probing chain length.
The same effect may apply if we compare \ac{wc} BL against cuDF but is visible at even lower multiplicities due to cuDF's lower throughput.
BL (1) (default) suffers from more bucket allocations compared to BL (2).
However, this effect mitigates with growing key multiplicity.
Our experiments showed that a CG size of 16 is optimal for \ac{wc} BL insertion.

As for retrieval (Figure \ref{fig:eval_multi_value_retrieve}), cuDF shows similar behaviour as during insertion.
With higher key multiplicities, performance decreases gradually.
In contrast, \ac{wc} \ac{oa} shows a nearly constant throughput between $0.66$-$0.72$ billion operations per second, which highlights the two benefits of our proposed \ac{cops} compared to cuDF's \ac{lp} approach.
(1) \ac{dh} ensures an overall shorter required chain of probings compared to \ac{lp}.
(2) Probing with a \ac{cg} allows for parallel retrieval of multiple values associated to the same key inside the same inner probing window.
cuDF uses a single thread per query which iterates over its probing sequence sequentially.
Furthermore, both \ac{wc} BL variants show nearly identical performance 
and consistently outperform \ac{wc} \ac{oa}.
Note that the overall retrieval throughput of our multi-value scenario is considerably lower compared to the retrieval step in the single-value benchmark due to two reasons.
(1) With increasing key multiplicity, the number of unsuccessful queries in the input set also increases, implying that more \ac{cg}s are executed than needed.
(2) Before the actual retrieval step, we have to calculate the value offsets for each key in the output array. 
For both \ac{wc} \ac{oa} and \ac{wc} BL a CG size of 4 shows optimal performance throughout all retrieval experiments.

\subsection{Use Case: Metagenomics} \label{sec:experiments:usecases:metacache}

The cost of DNA sequencing has decreased exponentially over the last years, making genomic data more accessible.
A common paradigm in bioinformatics is to store and index sequence data as sets of $k$-length substrings (called $k$-mers).
We have thus explored the efficiency of \ac{wc} for indexing large amounts of genomes for metagenomic classification tasks in comparison to the popular CPU-based tools Kraken2~\cite{Kraken2} and MetaCache~\cite{metacache} -- both using hash tables as their primary index data structure -- as a case study.
Note that such hash tables can be used for a variety of other bioinformatics applications, too.

We store $k$-mers as keys along with their corresponding genomic meta-information as values.
Metagenomic sequencing reads are classified by querying their own set of $k$-mers against the constructed hash table and subsequently evaluating the returned values.
$k$-mer reference database construction is typically the most time consuming part and can take several hours.
In this work, we focus on the MetaCache approach, which employs an efficient subsampling technique based on minhashing \cite{broder2000identifying} in order to reduce the overall amount of to-be-stored $k$-mers, with minimal loss in terms of classification accuracy.
In order to alleviate the mentioned bottleneck during the construction of the reference database, we chose to port parts of MetaCache's purely CPU-based construction phase to the GPU by utilizing GPU hash table building blocks provided by \ac{wc}.
To increase overall throughput, we also ported the $k$-mer generation and minhashing step to CUDA.
Using a single CUDA kernel, we process the sample sequences on the GPU in a data-parallel fashion and insert the resulting $k$-mers into a multi-value hash table provided by our proposed \ac{wc} library by using its device-sided interface.

\begin{figure}[t]
    \centering
    \includegraphics[width=0.85\linewidth]{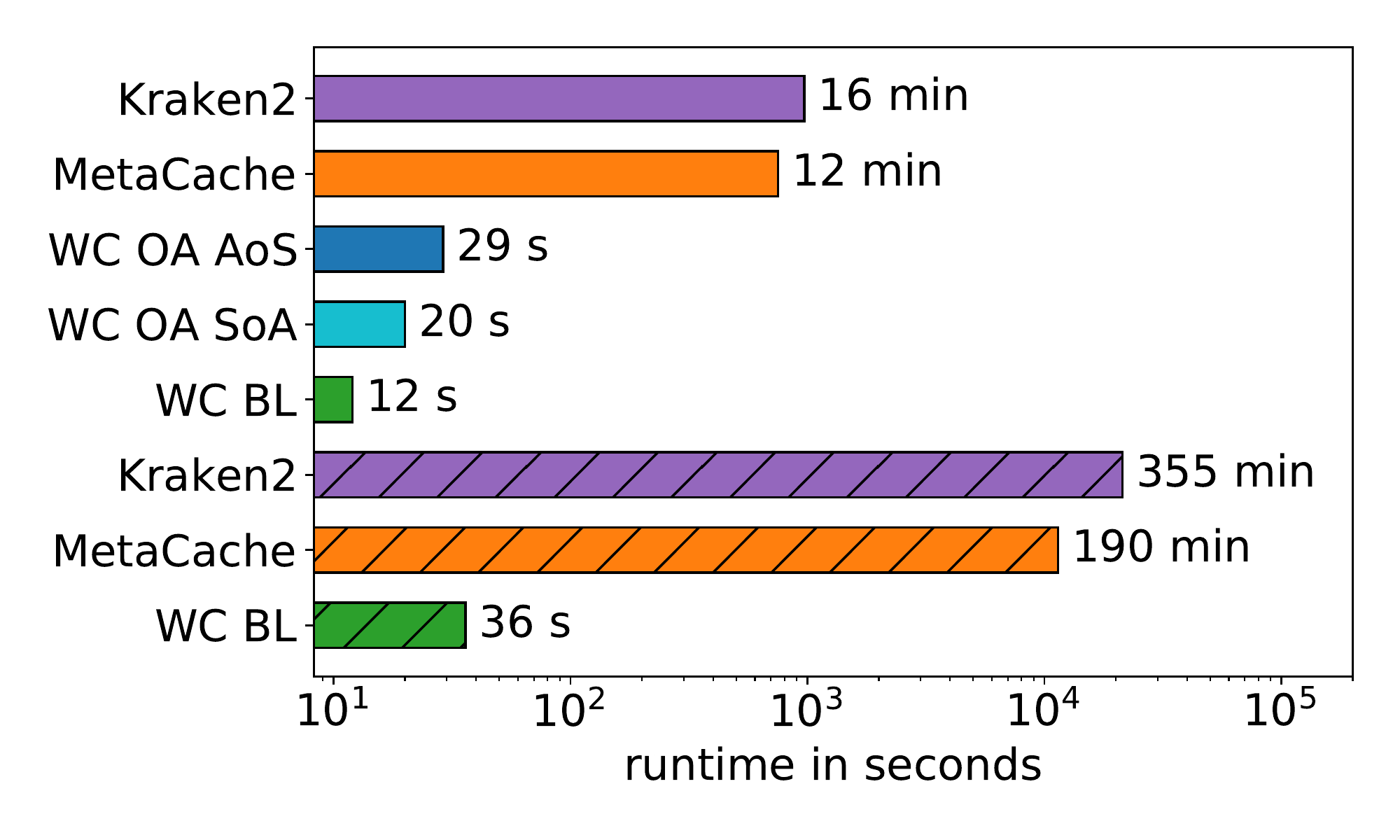}
    \caption[Metagenomics]{Comparison of metagenomic database construction times for small (solid bars) and large (hatched bars) datasets. 
    Small-scale construction on a single GPU compares different \ac{wc} multi-value hash table variants.
    Large-scale construction uses \ac{wc}'s bucket list on 8 GPUs.}
	\label{fig:metacache}
\end{figure}

First, we tested which of \ac{wc}'s multi-value implementations was best suited for metagenomic database construction by building a single-GPU hash table.
We therefore limited the overall hash table size to 28 GB and used a single GV100 GPU of System 1 to build a database for 18 GB of bacterial reference genomes.
The remaining 4 GB (of the 32 GB device memory) could then be used for batched input processing and retrieval.
Our implementation uses one CPU thread for extracting sequences from input files and another thread managing a double buffer for batched data transfer to GPU and insertion. 
The results for \ac{wc}'s variants compared to Kraken2's and MetaCache's default CPU construction are shown in Figure \ref{fig:metacache}.
Note that the key distribution for this data set is highly skewed.
Although the average number of values per key is near $11$, about a third of the keys has only one associated value and a small amount of keys occurs hundreds of times, which benefits the dynamic growth strategy of \ac{wc}'s bucket list (BL) hash table.
Overall, \ac{wc} BL achieves a speedup of 80.7 and 62.5 compared to Kraken2 and MetaCache, respectively.  



Because the memory of a single GPU is too small to hold large-scale reference genome databases, we also explored the usage of multiple GPUs.
For this test we used a reference genome dataset intended for food sequencing \cite{metacacheAFS}, which consists of bacterial, viral, and archaeal as well as animal and plant genomes.
About 120 GB of genomes were used to build a distributed database using the 8 GPUs of System 2 in parallel.
Figure \ref{fig:metacache} also shows the large-scale runtimes for Kraken2 and MetaCache compared to \ac{wc}'s bucket list (BL) hash table, which turned out fastest in the single-GPU benchmark.
Building the multi-GPU database took 36 seconds resulting in a speedup of 592 and 317 compared to Kraken2 and MetaCache, respectively.

\section{Conclusion}\label{sec:conclusion}

Rapidly growing data volumes in many fields such as bioinformatics have led to an increasing demand for fast associative data structures on modern parallel architectures. 
State-of-the-art GPU-based solutions are either only applicable to a small range of practical use cases or show unsatisfactory performance characteristics and storage densities.
A prominent example of the latter is that many hash map implementations require trading off runtime performance for memory efficiency because their throughput decreases significantly for high load factors.
Throughput of the few existing GPU multi-value hash maps decreases dramatically for key distributions with many associated values per key.

We have presented massively parallel hashing data structures and associated algorithms for single-value and multi-value hash maps that can be adapted to a variety of use cases. Their customization within a library (\ac{wc}) is achieved through a set of fundamental building blocks for data layout abstractions.
We exploit the fast memory interface of modern GPUs by means of a parallel probing scheme based on CUDA cooperative groups where threads communicate using fast collective operations such as group votes.  

We have demonstrated that \ac{wc} outperforms other state-of-the-art solutions by achieving billions of table operations per second on a single GPU even under very high load factors. 
Both our multi-value hash maps (pure \ac{oa} and bucket list hash maps) provide robust throughput over a wide range of possible key multiplicities significantly outperforming NVIDIA RAPIDS cuDF, especially for bulk retrieval operations.
We have further shown how to scale hash tables to multiple GPUs with fast NVLink interconnect in order to overcome the memory limitations of a single GPU. Using our library, we were able to accelerate a real-world bioinformatics application - metagenomic classification - on both single GPUs as well as on a multi-GPU DGX server. 

Scaling to even bigger datasets could be achieved by extending our library to GPU clusters. While \ac{wc} is specifically designed for GPUs, the introduced concepts could serve as a basis for efficient implementations on other accelerators such as modern FPGAs.

\ac{wc} is written in C++/CUDA-C and will be made publicly available upon the acceptance of the paper.

\section*{Acknowledgment}
Parts of this research were conducted using the supercomputer Mogon II and/or advisory services offered by Johannes Gutenberg University Mainz (hpc.uni-mainz.de) which is a member of the AHRP and the Gauss Alliance e.V.

\bibliography{references}
\bibliographystyle{IEEEtran}

\end{document}